\documentclass{iopconfser}
\usepackage{makecell}
\usepackage{graphicx}
\usepackage[skip=0.2\baselineskip]{caption}
\usepackage{hyperref}
\begin{document}

	\title{Pinpoint resource allocation for GPU batch applications}
	
	\author{Tim Voigtländer$^{1}$, Manuel Giffels$^{1}$, Günter Quast$^{1}$, Matthias Schnepf$^{1}$ and Roger Wolf$^{1}$}
	
	\affil{$^1$Karlsruhe Institute of Technology, Karlsruhe, Germany}
	
	\email{ tim.voigtlaender@kit.edu, Manuel.Giffels@kit.edu, guenter.quast@kit.edu, matthias.schnepf@kit.edu, roger.wolf@kit.edu}
	
	\begin{abstract}
		With the increasing usage of Machine Learning (ML) in High energy physics (HEP), there is a variety of new analyses with a large spread in compute resource requirements, especially when it comes to GPU resources. For institutes, like the Karlsruhe Institute of Technology (KIT), that provide GPU compute resources to HEP via their batch systems or the Grid, a high throughput, as well as energy efficient usage of their systems is essential. With low intensity GPU analyses specifically, inefficiencies are created by the standard scheduling, as resources are over-assigned to such workflows. An approach that is flexible enough to cover the entire spectrum, from multi-process per GPU, to multi-GPU per process, is necessary. As a follow-up to the techniques presented at ACAT 2022, this time we study NVIDIA's Multi-Process Service (MPS), its ability to securely distribute device memory and its interplay with the KIT HTCondor batch system. A number of ML applications were benchmarked using this approach to illustrate the performance implications in terms of throughput and energy efficiency.
	\end{abstract}
	
	\section{Usage of GPU resources by the CMS Collaboration}
	High energy physics (HEP) has become a compute resource intensive field of research due to the immense amounts of data provided by the CERN LHC \cite{LHC} and its experiments. This trend is sure to continue as the LHC enters its high luminosity phase in the coming years and the available data increases by close to an order of magnitude. The drive towards efficient use of the necessary, but costly hardware has been a topic of increasing importance for all related experiments, including the Compact Muon Solenoid (CMS) Collaboration \cite{CMS}. The use of GPUs has been a topic of great importance for this goal as CMS estimates that GPU resources can provide up to 2.8 times the amount of compute capacity for the same price when compared to CPU hardware, thereby leading to a 20-26\% increase of available compute resources by 2030 \cite{GPU_estimate}. In addition, many machine learning based analyses require the massive parallel processing power provided by GPU resources to be feasible at all. One GPU cluster, that has been set up for the purpose of supporting CMS and local HEP analyses, is the Throughput
	Optimized Analysis System (TOpAS) \cite{TOpAS}. It is a throughput oriented compute cluster with 24 V100S, 8 V100 \cite{V100} and 24 A100 \cite{A100} NVIDIA GPUs, providing a significant amount of parallel processing power to end user analyses of all shapes and sizes. 
	
	This paper is a follow-up to a 2022 ACAT contribution \cite{ACAT2022} and builds on the presented topics while providing an alternative to the NVIDIA Multi-Instance GPU (MIG) \cite{MIG} setup discussed at that time. The main feature that is benchmarked here is NVIDIA's Multi-Process Service (MPS) \cite{MPS}, a software that supports the execution of concurrent CUDA \cite{cuda} processes on a GPU. In addition, a comparison with the previously covered MIG is drawn and the use of the MPS software as part of the TOpAS batch system is illustrated.
	
	\section{Resource granularity in the context of GPU}
    To keep operating costs as low as possible, one of the most important goals of HEP-computing is the maximal utilization of its deployed hardware. The drive for maximal utilization applies to common resources like CPUs and memory, but even more so for costly hardware like GPUs.
	
	In many cases, a single workload will not fully utilize a hardware resource and multiple different tasks (in this context also referred to as jobs) have to run concurrent on the same hardware to achieve optimal utilization. Such sharing is inherently risky, as it is uncertain to which degree jobs influence each other or if there are enough resources to run all of them at once. One solution to some of these issues can be found in the various implementations of batch systems. 
	
	The implementation of choice for TOpAS, HTCondor \cite{HTCondor}, is commonly used in HEP computing for CMS because of its adaptability and open source nature. It provides resource provision, job scheduling and limited isolation between the jobs. The jobs that are sent to match with the available job slots through such a system can have widely varying hardware requirements, depending on their tasks. Especially with GPU-jobs submitted by end users for their analyses, this can range anywhere from a few percentages of a single GPU's processing power to the use of multiple GPUs at once. While isolation is important, a movable boundary between the job slots is necessary to properly fit as much work as possible on the hardware. For some hardware like RAM or scratch space, this is fairly easy, as they can be distributed in pieces of almost any size. CPU cores are more coarsely granular, and real life clusters often do not assign specific CPU cores to jobs, but instead only keep track of the overall CPU usage across all available cores. 
	
	GPUs would profit from a similar mechanism, but two reasons make this challenging. Each GPU possesses an amount of device memory (VRAM) that is used to store data while it is worked on by the GPU. This need for data locality makes it difficult to treat multiple GPUs uniformly, as one would do for CPUs, leading to them being individually assigned to their respective job(s). In addition, performance deteriorates with multiple processes running on the same GPU without additional supporting software or hardware features. Due to these complications, the default of HTCondor is that one GPU can only be assigned to at most one job, although it is possible to assign multiple GPUs to a single job. One solution to this issue of coarse granularity of GPU resources has been the topic of the previous contribution \cite{ACAT2022} and this paper builds on those findings by presenting an alternative.
	
	\section{NVIDIA MPS}
	\label{chap:MPS}
	While the previous work has concentrated on comparing MIG with the default case, this paper concentrates on the more software based approach of MPS. The MPS software exists and is supported in its current form since NVIDIA's \textit{Volta} GPU-generation. While a software with the same name was available for older graphics cards, it does not share the same features as with newer cards. We will focus on the post-\textit{Volta} capabilities of this service in this paper, but information on the pre-\textit{Volta} behavior can be found in the documentation \cite{MPS}. It should be noted that both MIG and MPS can be used together if necessary.
	
	The default behavior of post-\textit{Volta} NVIDIA GPUs, when confronted with multiple processes trying to access them, is to apply time slicing. Each of the processes opens a context for the GPU, but these contexts are not allowed to execute concurrently and are instead allotted time on the GPU in a round-robin scheduling. While such usage is safe, it also leads to severe under utilization of resources, especially when a larger number of small scale contexts is involved, since only one of them can work at a given time. Tests with OpenCL \cite{OpenCL} on NVIDIA V100 and A100 GPUs have indicated that this is also the default behavior for non-CUDA processes working with the GPU. The previously discussed MIG approach attempts to solve this issue by splitting the GPU into multiple smaller parts, each of which acts as an independent GPU, leading to multiple concurrent contexts when viewing the GPU as a whole.
	
\begin{figure}[h]
		\begin{minipage}{0.5\textwidth}
			\includegraphics[width=\textwidth]{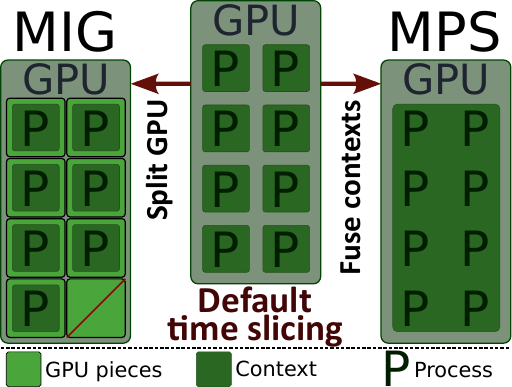}
			\caption{\label{sketch}Sketch of MIG and MPS behavior compared to default. Time slicing is only used with multiple contexts per GPU/GPU-piece. MIG reserves one piece for overhead.}
		\end{minipage} \hspace{0.05\textwidth}%
		\begin{minipage}{0.45\textwidth}
			\includegraphics[width=\textwidth]{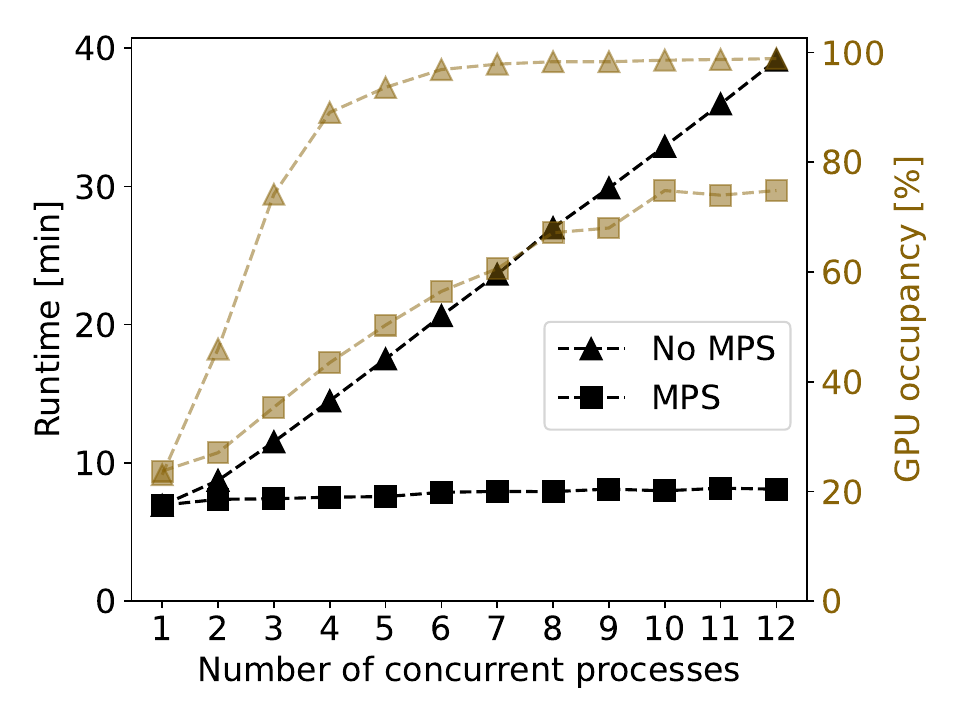}
			\caption{\label{FigMPSBare}Runtime and GPU occupancy of bare-metal workload with and without MPS.}
		\end{minipage} 
	\end{figure}

	MPS goes the opposite direction by taking multiple contexts and fusing them into one larger context. This also results in the intended \textit{one context per GPU}, as with the MIG approach. A sketch of the default, the MIG and the MPS use case is illustrated in figure \ref{sketch}. MPS provides fully isolated GPU address spaces for VRAM protection and limited error containment. This ensures that if any of the contexts combined by MPS experience a fatal error that affects the entire GPU, all processes generate the error, protecting them from incorrect results. It also supports limited execution resource provisioning for both, active thread usage and pinned device memory, enabling protection regarding overbooking of device resources. These features are helpful in isolating processes from each other, but they are not as effective as the ones provided through MIG, as will be discussed in section \ref{chap:compare}. 
	
	The setup for MPS is very flexible, as it optimizes any CUDA context that has access to it, enabling it to work with contexts of widely varying sizes at the same time. This makes it well suited for the use with the variety of different workloads that are sent to a batch system, as no reconfiguration is necessary compared to MIG. The theoretical maximum impact would be achieved with up to 48 processes sharing a single GPU efficiently, although it is unlikely that this many GPU processes would still fit the overall resource limits of the hardware.
	
	\subsection{Details on special MPS behavior}
	There are a number of behavior details that are not obvious from the documentation alone, which will be touched upon in the following, as they are of importance for the later introduction of MPS into our local HTCondor setup.
	
	All processes that MPS fuses have to be from the same user. Processes from different users will never run in the same MPS process on the same device, instead they will wait for the previous user's processes to finish before switching the \textit{owner} of the MPS process to the new user. As all jobs run as a common user on TOpAS, instead of themselves, this is not an issue. As a result, the setup presented in the following can be used by an entire group that submits jobs to such a batch system without any switching of the active MPS owner.
	
	By default, one MPS process will be spawned for the entire machine, independent of how many GPUs are installed. For large numbers of client processes that have to be managed by MPS, this can lead to performance issues as the MPS process might be overwhelmed. The MPS documentation mentions a limit of 48 client CUDA contexts per device. Tests with a machine with multiple GPUs have shown that slowdown starts to occur around 50 processes managed by one server. This does align with the stated 48 processes, but it indicates that the limit is not just for the number of contexts per GPU, but also for the number of allowed contexts per MPS process. Setting up an MPS process for each GPU individually lead to improved performance when compared to a single MPS process for the entire machine.
	
	Any context that is not fused by an MPS process will be placed into the time-sliced scheduler together with the MPS context. It is therefore possible to have multiple MPS processes run on the same GPU, although it is not recommended for performance reasons. It also allows to explicitly exclude certain contexts from an MPS process, which is necessary for lightweight monitoring processes that run as non-default users. Normally, they would have to wait for all other work of the MPS process to finish before running a simple query, but in this setup, they can run interleaved through time slicing with the main MPS processes, leading to marginal impacts on the overall performance.	
	
	\section{Viability check and performance benchmarks}
	In order to test whether MPS can provide performance improvements to a HEP compute cluster, a set of benchmarks based on the ones from the 2022 contribution \cite{ACAT2022} were performed. Both, the workloads and the hardware were the same as before. A set of small scale TensorFlow \cite{Tensorflow} neural network training was run on a machine with eight NVIDIA A100 GPUs. The workload is part of a real HEP workflow and consists of a total of 414 jobs with slightly differing input parameters. For the first two benchmarks, the same single job as in the previous contribution was used. For the final benchmark, all jobs were used.	For additional details, we refer to the ACAT 2022 contribution \cite{ACAT2022}.
	
	\subsection{Bare-metal performance benchmark}
	In the first benchmark, the same workload with a fixed number of training epochs was run on bare metal, once with MPS activated and once without. In all runs, processes were also assigned two CPU threads each. To see the direct impact on concurrent processes, multiple runs with differing numbers of instances of the same training were run on one A100 GPU. The maximum number of concurrent copies was limited to a maximum of twelve due to VRAM constraints in this benchmark for both versions. The results can be seen in figure \ref{FigMPSBare}. It can be observed, that the runtime without MPS increases close to linearly as more concurrent processes are added to the GPU. The reported GPU occupancy also increases sharply, without visibly benefiting the runtime of the workload, indicating the expected time slicing behavior discussed in section \ref{chap:MPS}. For the benchmark with MPS, the runtime shows little to no increase and the GPU occupancy grows much more gradually with the increasing number of processes. This indicates that time slicing was not applied here, and instead the different processes perform their work on the GPU concurrently. The overall increase in throughput measured for twelve concurrent processes is close to a factor of five.
	
	\subsection{Energy efficiency benchmark}
	\label{chap:energy}
	The second benchmark intends to judge the energy efficiency of different methods, described as the amount of energy necessary to run the training process of the workload for a fixed number of epochs. The benchmark scenarios aim to simulate the optimal high throughput case in which as many of the available hardware resources as possible are busy, as this is also the point at which the hardware runs the most energy efficient. Apart from the CPU only scenario, none of the scenarios manage to completely utilize the available CPU resources. While this is not ideal, it was not possible to fully utilize CPU and GPU at the same time during the GPU scenarios. It is therefore not possible to give a perfectly fair comparison between the methods, as additional work could have been performed on the remaining CPU threads with unknown impact on the overall power consumption. The tools used to measure the power usage during the benchmark (\textit{ipmitool} \cite{ipmi} for overall power consumption and \textit{nvidia-smi} \cite{nvidiasmi} for GPU power consumption only) state a systematic uncertainty of 5 \%, which leaves us with a somewhat uncertain but still meaningful result.
	
	There are five setups that have been considered. Firstly, the CPU only setup (CPU), in which each training was assigned four CPU threads and a total workload of 64 instances were running concurrently to occupy the 256 available CPU threads of the machine. The choice for this was determined by a set of earlier test runs, that resulted in four being the most energy efficient number of CPU threads for that specific workload. The idle power use of the GPUs has been removed from the measurements for this scenario to emulate a machine that is purely CPU based. The other four setups try to utilize the GPUs in different ways, relevant for the use in a batch system. The second setup (Default) simulates the exclusive use that is the default for the use of GPUs with HTCondor. It assigns only one job to each GPU at most, generally resulting in only one process running on that GPU. The third (Simple sharing) is an example of a GPU that is allowed to run multiple jobs at once without any additional supporting setups like MIG or MPS. As in the first benchmark, up to twelve instances of the workload fit onto the GPU at the same time before it runs out of VRAM. The fourth scenario (MPS) is the same as the third, with the exception, that MPS is also enabled to better support the concurrent execution. The fifth and final setup (MIG) is a rerun of the one from the 2022 contribution, in which each GPU was split into the maximum number of independent GPU parts. As a result, a maximum of seven workload instances can be run concurrently per GPU. The results of the benchmark, including run time of the individual workloads, power usage, power efficiency and the number of concurrent benchmark instances on the machine, can be found in table \ref{tab:power_bench}.
	
	\begin{table}
		\centering
		\caption{Results of the energy efficiency benchmark. Scenarios as explained in section \ref{chap:energy}. The energy per training is calculated as the power draw multiplied with the average runtime of one instance divided by the number of concurrent instances.}
		\begin{tabular}{llllll}
			\Xhline{3\arrayrulewidth}
			Benchmark scenarios & CPU & Default & Simple sharing & MPS & MIG\\
			\hline
			\# of concurrent instances & 64 & 8 & 96 & 96 & 56\\
			Average runtime [min] & 82.07 & 10.56 & 41.43 & 16.23 & 9.41\\
			Total power draw [W] & 664.80 & 748.22 & 1202.58 & 1396.29 & 1327.55\\
			Energy per training [kJ]& 51.95 & 59.25 & 31.15 & 14.17 & 13.39\\
			\Xhline{3\arrayrulewidth}
		\end{tabular}
		\label{tab:power_bench}
	\end{table}
	
	It can be seen that the default setup is highly inefficient compared to all other setups, including the CPU one. All other GPU scenarios show better efficiency due to the higher number of concurrent instances when compared to the default case and the faster runtime when compared to the CPU case. The simple sharing scenario is less than half as efficient as both optimized cases with MPS or MIG.
	
	\subsection{HTCondor scaling benchmark}
	\label{scale}
	As a final benchmark before deployment, the same workload as the scaling benchmark from the 2022 contribution was repeated with additional benchmark scenarios. A total of 414 jobs with similar workloads to the ones of the above benchmarks were sent to run on the benchmark machine through an HTCondor batch system. The four scenarios examined here are the ones from the previous benchmark without the CPU case. It should be noted that the hardware resources are not as intensively utilized as in the bare metal benchmarks above. Some dead time between the end of one job exists before the next job starts and towards the tail end of the benchmark, fewer jobs are running at the same time. While the dead time between jobs is an expected feature when working with batch systems, the decreasing concurrency towards the end of the benchmark is not. A high throughput system is intended to run a consistent stream of jobs and as such, the batch system is not meant to run out of jobs in general. To take these discrepancies into consideration, a cut was applied to ensure that only the part of the benchmark is considered, during which the benchmark machine is fully utilized. The throughput shown in table \ref{tab:throughput_batch} is in the form of epochs per second to account for an inherent randomness in the number of epochs that are necessary for the individual trainings to converge.
	
	\begin{table}
		\centering
		\caption{Results of the HTCondor scaling benchmark. Scenarios as explained in section \ref{scale}. Total throughput given as epochs per time to account for a varying number of epochs per job.}
		\begin{tabular}{lllll}
			\Xhline{3\arrayrulewidth}
			Benchmark scenarios & Default & Simple sharing & MPS & MIG\\ \hline
			Total Throughput [Epochs/s] & 1.94 & 10.06 & 18.16 & 14.14\\
			\Xhline{3\arrayrulewidth}
		\end{tabular}
		\label{tab:throughput_batch}
	\end{table}
	
	It can be observed that the default case, with only one job allowed per GPU, has the least throughput. The approach to share a GPU without actually deploying any supporting software or hardware features is a significant improvement. The best performance is shown by the MIG and to an even greater degree the MPS setup. The main difference in the performance between the benchmark scenarios is the number of available job slots. The throughput for an individual job is nearly equal between all setups except for the \textit{simple sharing} one, as that one is bottle-necked by the time slicing behavior explained in section \ref{chap:MPS}.
	
	\section{Deployment in production for an HTCondor batch system}
	The setup with MPS has been rolled out for a number of machines on the TOpAS cluster. The machines in use are not the same as during the described tests, with the main difference being that the setup was applied to three machines that utilize NVIDIA V100S GPUs instead of the A100 GPUs for the benchmark. The setup consist of an MPS daemon process running in the background, an HTCondor configuration to manage the resource allocation and a script to manage the supervision through MPS for the running jobs. The variable \texttt{CUDA\_MPS\_PIPE\_DIRECTORY} of the MPS service is set to a non-default value in order to have MPS only affect processes that actively set this variable to the same value in their environments. All other processes that run on the GPU are unaffected by MPS and subject to the normal time slicing behavior. The configuration file sets up one partitionable slot and a VRAM pool for each of the eight physical GPUs. In addition, the existing GPUs are advertised eight times each and can therefore be assigned to up to eight different jobs. An incoming job with a requirement for a GPU and an amount of VRAM can be assigned the required resources from one of the partitionable slots. With this setup, jobs can be placed on an individual GPU as long as the required VRAM is available. This setup does not allow for multi-GPU jobs as it would require resources from multiple different partitionable slots, which is generally not supported. Finally, the script modifies the initial environmental settings to steer the supervising MPS service, including the \texttt{CUDA\_MPS\_PIPE\_DIRECTORY} variable. In addition, it sets the \texttt{CUDA\_MPS\_PINNED\_DEVICE\_MEM\_LIMIT} variable, preventing processes from occupying more VRAM than requested. 
	
	This setup has been in deployment since May 2024 with no major issues. During peak usage, between four and five jobs have been executed on each of the available physical GPUs concurrently. This is a considerable improvement to the default setup, where only one job is allowed per GPU. 
	
	\section{Comparison to MIG}
	\label{chap:compare}
	Both MIG and MPS can provide significant improvements to the throughput of GPU jobs in a batch system, but they also have their respective up- and downsides. MPS is available for a number of graphic cards that do not support MIG, such as the NVIDIA V100 GPUs that are part of TOpAS. MPS is generally easier to set up and allows for greater flexibility due to its software based nature. This has allowed it to be included as part of the TOpAS batch system without complex changes to the underlying HTCondor installation. The combination of multiple processes through MPS complicates the distinction between jobs for the purpose of monitoring, as only a single MPS process is attributed all the work that is performed on the GPU. With some extra effort, assigned VRAM can still be monitored, but GPU utilization can only be roughly estimated. Although MPS offers tools to allocate hardware resources and enforce limits, these boundaries can sometimes be breached, resulting in overutilization until the policy violation is detected. Additionally, problematic processes can affect other processes supervised by MPS if the error is sufficiently severe to affect the entire device. Finally, GPU processes spawned by non-CUDA libraries are not supported by MPS, leading to errors or unexpected behavior.
	
	Due to its hardware based nature, MIG is more difficult to set up and its flexibility is limited. In summary, this makes it difficult to use MIG dynamically in a batch system, as only the preset configuration of GPUs would be available. The size of those pre-configured GPU pieces might not fit the requirements of the requested jobs. On the other hand, this rigidity leads to great isolation between the individual GPU pieces and enables access to the monitoring data of each of them. CUDA-external GPU libraries show the same behavior as with full GPUs, as the supervision is not directly CUDA related, making it safer to use with those libraries than the CUDA based MPS.
	
	TOpAS provides a good testing ground for comparisons of this nature, and has lead to additional insights concerning the usage of different GPU models. The A100 GPUs are usually requested for larger scale applications that require the full VRAM available, while the smaller V100 GPUs tend to be occupied by small scale jobs that only require few GPU resources. As such, it is counterproductive to remove A100 GPU resources from the batch system to split them into smaller GPUs, as long as the V100 GPUs could take on those jobs. As MIG is not available for the V100 GPUs, the only way to improve their utilization is through MPS. The variety of workflows with a wide array of hardware requirements should be taken into consideration when acquiring hardware for future HEP compute cluster projects.
	
	Over all, MIG would be the more suitable solution in many cases due to the higher degree of security, but the requirement of newer hardware limits the technology significantly. Even for hardware that is capable of using it, the complex and rigid setup currently makes it difficult to use in a batch system environment.
	
	\section{Conclusion}
	The Multi-Process Service (MPS) allows for multiple CUDA contexts to be fused into a single one. Due to the limitations of the default time slicing scheduling, this allows for greater performance when compared with only placing multiple contexts on the same NVIDIA GPU.
	
	The performance of MPS was benchmarked on bare metal and in the context of an HTCondor batch system, using the TOpAS cluster. The viability of its use as part of a batch system was tested and the benefits and risks have been described. The benchmarks have shown that in an ideal scenario, the addition of MPS can lead to speedups of close to five when compared to the non MPS scenario. The benchmark scenario with MPS has shown itself to be more than twice as efficient as the one without MPS, and closely behind the MIG approach. When tested in the context of a real batch system, MPS has shown a speedup over the current default usage of more than nine, although a simpler approach of sharing that does not involve MPS has also shown a speedup of five over the default non-sharing approach. In addition, the findings from the 2022 ACAT contribution were reproduced.
	
	The introduction of MPS into the TOpAS batch system has led to an improved throughput of four to five for those nodes during peak usage, and the setup has been described. The technologies MPS and MIG have been compared and the topics of performance, security, ease of use and flexibility have been covered.
	\newpage
	\bibliography{sources}

\end{document}